\documentclass[showpac,nofootinbib]{revtex4}
\usepackage{graphicx}
\begin{document}
\title{DNA ejection from bacteriophage: towards a general behavior for osmotic suppression experiments}
\author{M. Castelnovo}\thanks{Laboratoire Joliot-Curie et Laboratoire de Physique -- Ecole Normale Sup\'erieure de Lyon, 46 All\'ee d'Italie, 69364 LYON CEDEX 07, FRANCE} 
\author{ A. Evilevitch}
\thanks{Department of Biochemistry, Center of Chemistry and Chemical Engineering, Lund University, P.O. Box 124, S-221 00 LUND, SWEDEN }
\date{\today}
\begin{abstract}
We present in this work \textit{in vitro} measurements of the force ejecting DNA from two distinct bacteriophages (T5 and $\lambda$) using the osmotic ejection suppression technique. Our datas are analyzed by revisiting the current theories of DNA packaging in spherical capsids. In particular we show that a simplified analytical model based on bending considerations only is able to account quantitatively for the experimental findings. Physical and biological consequences are discussed.
\end{abstract}

\maketitle

\section{Introduction}
\label{intro}
Viruses have developped various specific strategies over the evolution in order to infect higher organisms. In the case of bacteriophages, it has long been realized that high elastic energy of viral DNA confined inside the capsid is passively used to inject DNA in the bacteria cell cytoplasm, at least in the early stage of infection (see for example references in \cite{riemer_78,molineux_06}). Over the last decade, there has been a renewed interest into both measuring and modeling the energetics of DNA packaged inside bacteriophage \cite{lambert_00}. 

On the experimental side, three different phages were studied by three different techniques: single molecule technique monitoring the packaging kinetics of $\phi 29$ \cite{smith_01,chemla_06}, fluorescent staining and light scattering monitoring T5 and $\lambda$ ejection kinetics \cite{frutos_05a,frutos_05b,mangenot_05,lof_07}, and osmotic suppression of DNA ejection from $\lambda$ phage \cite{evilevitch_03,evilevitch_04,evilevitch_05,grayson_06,evilevitch_06}. Only this last method is thought to measure direct equilibrium properties of DNA packaged inside capsid for various length \cite{evilevitch_04,grayson_06}.

On the theoretical side, it has been early understood that the main energetic balance involves bending energy, due to the small size of capsid relative to DNA persistence length, and short-range interstrand interactions due to relatively high volume fraction of DNA inside capsid \cite{riemer_78,bloomfield_91,odijk_98}. The former energetic contribution is described using classical linear elasticity theory \cite{odijk_98}, while the latter is thought to involve subtle balance between electrostatic and hydration interactions \cite{kindt_01,tzlil_03,purohit_03a,odijk_03}. Direct comparison of experimental results obtained by the osmotic suppression technique with theory was performed in references \cite{evilevitch_04,grayson_06} for $\lambda$ phage, and quantitative agreement was found. However, no other phages were characterized so far using this technique, therefore raising the question of the bearing to other phages of the conclusions drawn from experimental and as well as theoretical study of $\lambda$ phage. It is the purpose of the present work to cure this lack.

T5 phage is a good candidate for extending osmotic suppression studies for several reasons. First, its capsid is among the larger of phages, and therefore the scaling of ejection force amplitude with capsid size and DNA content can be addressed experimentally for the first time. Second, kinetics experiments on T5 have shown interesting stepwise features in the passive ejection process and the influence of these steps for equilibrium measurements is not known \cite{frutos_05a}. Finally the main membrane protein receptor FhuA for T5 can be purified and solubilized as described in \cite{boulanger_96}, which makes osmotic suppression measurements possible. The comparison of ejection force obtained by the same experimental method for two markedly different phages will therefore allow to validate the bearing of this technique.
Similarly, the quantitative interpretation of datas using a single model is a crucial test for validating the theory.

The osmotic suppression technique is based on the observation that \textit{in vitro} DNA ejection from bacteriophage $\lambda$ is partially or fully suppressed by bulk osmotic pressure of a polymer solution (poly-ethylene glycol, or PEG). These experiments, combined with appropriate modeling of osmotic pressure effect, allow to infer the magnitude of ejecting force \cite{castelnovo_03,evilevitch_04}. This force increases as more DNA is present in the capsid, and reaches a tens of $pN$ for fully packaged infectious wild-type $\lambda$ phage (genome length$=48,5kbp$).  Additional osmotic suppression experiments showed that ejecting force is shown to depend on the composition of the buffer as well \cite{evilevitch_07}. 

In our attempt to find the most appropriate model describing our experimental datas, we realized that most of the realistic theories of DNA packaging inside closed volume roughly fall into two large classes: models describing \textit{structureless} semiflexible rod packaging with finite thickness, therefore ignoring the peculiarities of DNA (charged polymer surrounded by discrete ions and solvent) \cite{marenduzzo_03,klug_03,lamarque_03,spakowitz_05,muthukumar_06,ali_06,locker_06,petrov_07}, and \textit{structured} models taking into account energetics of DNA-DNA interactions in a more accurate way \cite{odijk_98,kindt_01,tzlil_03,purohit_03a}. The former class of model focuses mainly on the way the conformation of DNA observed in capsids is achieved, while the latter is more interested in the energetics at fixed inverse spool conformation. Surprisingly, these two different type of models reproduces the ejection force datas of Smith \textit{et al.} on $\phi 29$ with similar accuracy. This leads us to revisit the model in order to better understand the reason of such an agreement and therefore gap the two kind of descriptions. The underlying goal is to find the relevant variable or scaling parameters to describe physics of DNA packaging in real phages.

This paper is organized as follows. In the next section, we present briefly the osmotic suppression experiments. Then some general considerations on DNA packaging inside capsid are provided. This leads us to propose a simplified model in section \ref{model}. The results of this theoretical approach together with experimental datas are discussed in section \ref{results}. Finally, conclusions of the present study are provided.

\section{Experiments}
\label{experiments}In this section, we describe briefly the osmotic suppression technique as applied to T5. The method and experimental results for $\lambda$ phage have been published elsewhere \cite{evilevitch_03}.

Bacteriophage T5 \textit{st(0)} with genome length 113 kbp (7.2$\%$ shorter genome compared to the wild-type T5 with 121.75kbp genome) was mixed with its purified recepetor protein, FhuA (solubilized in 0.03$\%$ w/v of LDAO surfactant), in order to allow phage to eject DNA in vitro. T5 phage and FhuA were kindly provided by Lucienne Lettellier and Marta de Frutos (respectively from Institut de Biochimie et Biophysique Mol\'eculaire et Cellulaire and Laboratoire de Physique des Solides, Universit\'e de Paris-Sud, Orsay, France). pH and ionic strength in the phage-receptor solution were set by adding 10mM trisHCl, pH 7.4, 100 mM NaCl, 1mM CaCl$_2$ and 1mM MgSO$_4$. The osmotic pressure resisting DNA ejection was set by addition of PEG8000 (MW 8000 g/mol) to the host solution. The relation between the osmotic pressure, PEG8000 concentration and temperature is well empirically established \cite{parsegian_86}, which allows us to determine the fraction of DNA ejected as a function of external osmotic pressure.

\begin{center}
\begin{figure}
\includegraphics[scale=0.5]{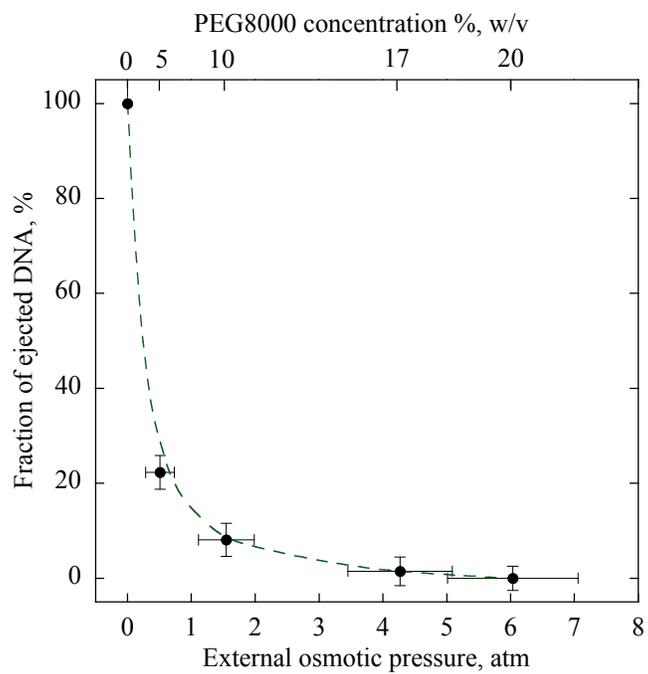}
\caption{\label{figure1}Osmotic suppression datas for T5 with error bars: fraction of ejected DNA \textit{vs} PEG8000 concentration (top axis) or osmotic pressure (bottom axis) for DNA ejection from bacteriophage T5 at 37$^\circ$C. The broken line is drawn to guide the eye. Horizontal and vertical error bars are derived from error propagation of deviations in PEG concentration and UV absorbance.}
\end{figure}
\end{center}

First T5 phage (at concentration $10^{12}$ virions/ml) was mixed with FhuA receptor at 1:200 phage to receptor ratio and DNase I was added to a final concentration of 20 $\mu$g/ml. The mixing was made at 0$^\circ$C for 30 minutes to allow phage to adsorb to FhuA receptor without ejection taking place at this temperature (which we checked by UV absorbance measurements). Then, the solution was mixed with PEG8000 at 0$^\circ$C and incubated for 1h to allow for complete mixing, followed by 3h incubation at 37$^\circ$C in a water bath to trigger the ejection of DNA from phage. After 3h incubation, the ejection has ceased and all ejected DNA is digested by DNase I. Without PEG the ejection from T5 was complete at 37$^\circ$C (which was also checked by UV absorbance knowing the intial concentration of phage particles). Phage capsids and their unejected DNA were then removed by ultra-centrifugation at 90000rpm (TLA-100 rotor) for 1h at 4$^\circ$C. Non-sedimenting DNA nucleotides of ejected DNA remain in the supernatant. The concentrations of ejected DNA are determined by UV absorbance measurement at 260 nm wavelength. Since there is also some external DNA always present in all phage samples, that is accessible to DNase I even without FhuA, we also need to account for this contaminant ``background'' DNA concentration. The concentration of this background DNA is determined in the same way with UV by taking the sample of phage with DNase I but without FhuA. The  fraction of ejected DNA as function of external osmotic pressure (set by PEG8000 concentration) is then determined by 
\begin{equation}
\% DNA_{ejected}=\frac{Abs(\mathrm{phage+FhuA+PEG+DNase\,I})-Abs(\mathrm{phage+DNase\,I})}{Abs(\mathrm{phage+FhuA+DNase\,I})-Abs(\mathrm{phage+DNase\,I})}
\end{equation}
see figure \ref{figure1}. The vertical bars reflect differences between repeated measurements of UV absorbance for the same sample with all conditions kept constant. Detailed description of this experimental procedure is provided elsewhere \cite{evilevitch_03}. 

\section{General considerations on DNA packaging energetics}
\label{general}

Several experimental evidences on bacteriophages support the ``inverse spool'' organization of DNA inside capsid \cite{earnshaw_77,lepault_87,cerritelli_97,lander_06}: according to this model, the molecule is wound in the interior of the capsid in an ordered way, filling first layers close to the walls, then layers of decreasing curvature radius, leaving finally a cylindrical hole inside the capsid devoid of DNA. This explains the widespread inverse spool labeling found in the litterature. The local hexagonal ordering of DNA in this case allows to reach high packing density. Using this picture, several groups developped equilibrium energetic models in order to identify and to quantify relevant physical factors at work in DNA packaging, starting from the pioneering works of Bloomfield \textit{et al.} \cite{riemer_78,bloomfield_91}. The main energetic contributions are associated to bending cost of the semi-flexible DNA and short-range DNA-DNA interactions. 
These are closely related  to capsid size and DNA volume fraction inside phages. 

In the case of $\lambda$ phage, the internal capsid diameter is of order $55nm$ \cite{dokland_93}, \textit{i.e.} of the order of DNA persistence length $l_p\simeq 50nm$, which represents the typical length scale of spontaneous bending due to thermal fluctuations. Packaging of DNA lengths longer than the persistence length inside such a small volume will therefore cost a significant amount of mechanical work. As more DNA is filled inside the capsid, the finite thickness of the molecule constrain it to follow a path with decreasing curvature radius. We expect therefore the bending contribution to the force ejecting DNA out of bacteriophage to increase sharply with the DNA content inside the capsid.

Similarly, the nominal nucleic acid volume fraction for fully packaged wild-type $\lambda$-phage is $\phi\simeq 0.59$.
By nominal volume fraction, we mean the ratio between DNA volume in a straight conformation ($V_{DNA}^{straight}=N_{bp}\pi r_{DNA\perp}^2r_{DNA||}$ with $N_{bp}=48500bp$ and $r_{DNA\perp}=1nm$ and $r_{DNA||}=0.34nm$) and internal volume of its capsid ($V_{capsid}^{in}=\frac{4\pi R_0^3}{3}$ with $R_0=27.5nm$ \cite{dokland_93}). 
This implies small interaxial distances of the order $d_s=2.7nm$ in ideal inverse spool conformation. At this length scale, the energetics associated to the polyelectrolyte nature of the DNA, as well as the discreteness of solvent molecules and counterions is expected to be another relevant contribution to the total energy of packaged DNA. Neglecting in a first approach the influence of bending, the energetics associated to these short-range interactions should be very similar to the one probed by osmotic stress experiments on DNA condensed phases \cite{rau_92}. This is the spirit of the model proposed originally by Tzlil \textit{et al.} \cite{kindt_01,tzlil_03}: the interstrand DNA-DNA interaction was estimated using experimental datas on the osmotic pressure of an hexagonal array of DNA molecules under similar buffer conditions \cite{rau_92}. Depending on these conditions, one might have either repulsive or attractive interactions, the latter case being provided by the presence of multivalent ions in the buffer that are known to collapse unconstrained DNA in solution. Using DNA interstrand interactions parametrized this way, and classical elasticity theory, Tzlil \textit{et al.} derived for the first time a model to compute DNA ejection force from spherical bacteriophage as function of packaged length. Following this approach, Purohit \textit{et al.} \cite{purohit_03a,purohit_03b,purohit_05,grayson_06} extended further this model along various lines. In particular, they used different capsid geometries (spherical, cylindrical, sphero-cylindrical) and buffer conditions, they analyzed the effect of discreteness of DNA on the completion of layers with different curvature radii inside the capsid and finally predicted some features about the strength of capsids. The main points of this model are that it allows on the one hand to predict quantitatively the total force ejecting DNA out of bacteriophage, and on the other hand to estimate the relative contributions of the two main driving forces (DNA bending and DNA-DNA interstrand interaction). In particular, using the parameters that best describe the experimental results of osmotic suppression experiments on $\lambda$-phage mutants with different nucleic acid lengths \cite{grayson_06}, one reaches the following conclusion: the maximal contribution of interstrand interactions for fully packaged capsid is slightly less than $20\%$ of total ejecting force. This means that bending cost provides the leading order of magnitude for this force. This observation is partially rationalized by the steepness of interactions: $r_o=0.27nm$ for repulsive interactions ($r_o$ is the exponential decay length of such interactions, cf \cite{rau_92}), and $r_o=0.14nm$ for attractive interactions, as taken from \cite{purohit_05}. Different layers of nucleic acids have therefore to be relatively close in space in order to probe this steep compression potential. This is thought to happen only at high packing fraction. As a consequence, DNA packaging energetics inside fully packaged wild-type $\lambda$ virions is marginally influenced by compression of neighbouring layers ($20\%$ of total ejecting force). 

The inverse spool picture used for the interpretation of experimental datas has been recently challenged by simulation results \cite{lamarque_03,spakowitz_05,muthukumar_06,locker_06,petrov_07}. The net outcome of these studies is that DNA organization inside phages might not be as well ordered as it is assumed in the previous \textit{structured} models. In particular, different final conformations or structure might be reached for the same nucleic acid content of the phage. Since the well-ordered inverse spool conformation of DNA is one of the main underlying assumption in the structured models, it is quite surprising that osmotic suppression datas about $\lambda$ phage are nevertheless described quite accurately \cite{grayson_06}. A key towards the understanding of such a feature relies in the observation that \textit{structureless} models \cite{klug_03,marenduzzo_03}, are also describing experimental datas (about $\phi 29$) with similar accuracy.

Based on the previous analysis, our main theoretical proposal in the present work is the following: the nucleic acid content dependence of \textit{ensemble-averaged} ejecting force (as obtained by osmotic suppression technique) is mainly given by the bending penalty associated to the confinement of DNA inside the capsid, \textit{whatever its conformation} as long as it satisfies excluded volume contraint. Indeed, since the final conformation of DNA inside capsids is unlikely to be unique after packaging, experimental datas obtained by osmotic suppression technique are ensemble-averaged, \textit{i.e.} averaged over different final conformations of the virions present in the solution. Moreover, one might expect that at given loaded length or packing fraction, the conformation of DNA does not influence significantly the bending energy, 
 since the packing density together with excluded volume contraint dictates similar variation of curvature radius in order to fill the capsid. This is partially justified by simulation results of Forrey \textit{et al.} \cite{muthukumar_06} on the one hand, and of Locker \textit{et al.}\cite{locker_06} on the other hand, where in both cases bending energies of different simulation runs, and therefore different DNA dynamic conformations, are shown to be similar for a given loaded length within few percents. The very same simulations also show that the total ejecting force is fluctuating from one final conformation to the other. 

\begin{center}
\begin{figure}
\includegraphics[scale=0.5]{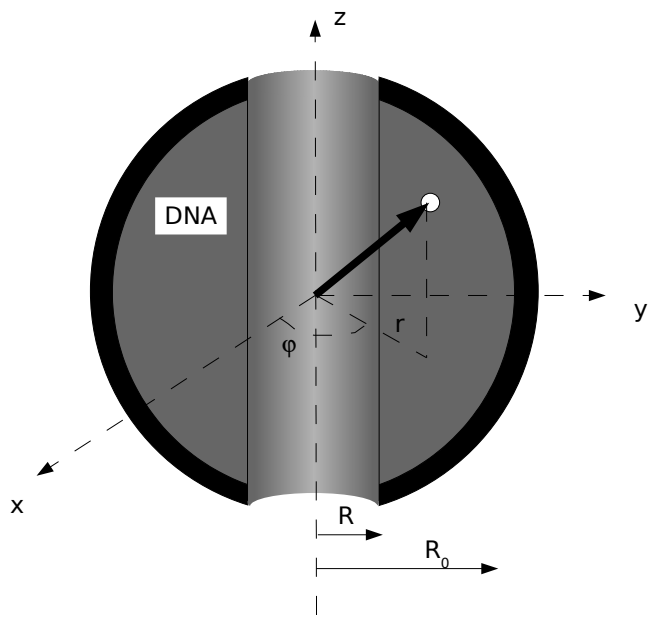}
\caption{\label{figure0} Uniform DNA condensate with cylindrical symetry used in the structureless model. Inner radius $R$ and outer radius $R_0$ of the condensate are shown.}
\end{figure}
\end{center}

In order to test the dominance of bending contribution with respect to interstrand interaction in the length dependence of ejecting force, we propose to calculate the former contribution in a structureless model of neutral semi-flexible polymer packaging described in the next section. Within this model, ensemble averaged DNA conformations inside the capsid are described by a cylindrical spool-like condensate of uniform DNA density (see figure \ref{figure0}). The molecular interactions are simply hard core repulsions, and are argued to contribute to the DNA density inside the condensate, independently of DNA packaged length. This last assumption is required in order to really test pure bending contribution, independently from DNA-DNA interactions. Notice that if the density depends on DNA packaged length, the packaging energetic probes the compressibility of DNA condensate, and therefore molecular interactions as well. The use of the bending energy contribution without further details of interstrand interactions is consistent with other structureless models \cite{marenduzzo_03,klug_03,muthukumar_06,locker_06}.

The underlying goal of the present proposal is twofold: first to provide a simple analytical formula describing osmotic suppression datas, and second to identify this way the relevant parameters describing the physics of DNA packaging in real phages. The results and bearing of such an analysis is illustrated in the next sections. 

Moreover, it is important to note that the DNA condensate with fixed uniform density is not inconsistent with the observed influence of buffer content on the amplitude of ejecting force \cite{evilevitch_04}. Indeed it has been shown experimentally that addition of multivalent ions in the buffer leads to a strong decrease of ejecting force amplitude. In the framework of our structureless model, this can be taken into account by different DNA densities associated to different buffer conditions. Indeed, it has been shown experimentally that density of DNA condensates in the presence of multivalent ions is strongly dependent on buffer conditions \cite{raspaud_05}. Note additionally that DNA persistence length is also known to depend on these conditions as well, as discussed for example in \cite{ariel_03}, providing therefore another source of ejecting force amplitude modulation by buffer conditions.

\section{Simplified model}
\label{model}
As it is suggested by the discussion in the previous section, our model of DNA packaging energetics inside bacteriophage will focus mainly on bending contribution. Indeed it can be shown that the contribution of excluded volume interactions within a uniform density condensate is simply a constant shift (independent of DNA packaged length) of ejection force, therefore justifying \textit{a posteriori} our focusing on bending energetics.\footnote{Within a structureless DNA condensate, the energy associated to excluded volume interactions of DNA monomers simply scales as $E_{int}\simeq \frac{kT}{2}\int_V d{\bf r}v \rho ({\bf r})^2$, where $v$ is the second virial coefficient of DNA monomers. Using the relation Eq.\ref{relation l rho} between DNA density and length toghether with constant density assumption, this energy is a linear function of DNA content $E_{int}\simeq\frac{kTv\rho_0}{2d_0}L_{in}$, and therefore it induces a constant shift in ejecting force $F_{int}\simeq \frac{kTv\rho_0}{2d_0}$. Since we are interested in the variation of ejecting force with respect to DNA packaged length, we focus only on bending contribution in the main text for the sake of concision.}. The capsid is assumed to be a sphere of inner radius $R_0$, which is roughly the case for T5 and $\lambda$ phages if one ignores the faceting of their capsids. Following Odijk \textit{et al.} \cite{odijk_03}, we use a continuous description of DNA inside the capsid, characterized by a coarse-grained monomer density $\rho({\bf r})$. This approach is well suited to describe ensemble-averaged disordered conformation of DNA, in contrast to recent theoretical works on DNA packaging assuming \textit{a priori} well-ordered conformation like the inverse spool conformation \cite{odijk_98,tzlil_03,purohit_03a}. The relation between DNA packaged length $L_{in}$ and monomer density $\rho({\bf r})$ is given by
\begin{equation}
\label{relation l rho}L_{in}=d_0\int_Vd{\bf r}\rho({\bf r})
\end{equation}
where $d_0$ is the monomer size ($d_0\simeq 2nm$ for DNA) and $V$ is the volume of the capsid interior. Similarly, the bending energy is written as
\begin{equation}
E_{bend}=\frac{kTl_pd_0}{2}\int_Vd{\bf r}\frac{\rho({\bf r})}{R_c({\bf r})^2}
\end{equation}
where $l_p$ is the DNA persistence length and $R_c({\bf r})$ the coarse-grained curvature radius of DNA at the location ${\bf r}$. At this step, three essential assumptions are needed to further develop the model: \textit{(i)} cylindrical symetry of DNA condensate inside the capsid; \textit{(ii)} concentric conformation of DNA inside the capsid; \textit{(iii)} constant density in the DNA condensate, that depends mainly on buffer conditions, but not on DNA packaged length.

The first assumption allows us to write the monomer density as $\rho(r,z,\varphi)=\rho(r,z)$, where the symetry axis ($z$-axis) is aligned with the portal protein(s) through which DNA is packaged/ejected. It is partially justified by both X-ray diffraction measurements and cryo-EM studies of particular phages \cite{earnshaw_77,lepault_87,cerritelli_97,lander_06}, although the precise organization of DNA within capsid is not known. In particular, this assumption does not imply any hexagonal ordering of part or the entire DNA condensate as in the ideal inverse spool structured models. The second assumption is strongly correlated to the first one. The average concentric conformation of the DNA allows us to specify the average curvature radius used to perform bending energy calculation, such that $R_c(r,z,\varphi)\sim r$. For the sake of simplicity, the prefactor is set to unity.  Finally the last assumption allows to introduce a constant monomer density $\rho_0$ in the DNA condensate, which is independent of DNA packaged length $L_{in}$. This might look like a very crude approximation,
 since curvature stress has been previously shown to crowd DNA inside the capsid within structured inverse spool model \cite{odijk_98}. However, since we want to highlight the role of bending energy in providing the most relevant contribution to \textit{ensemble-averaged} DNA ejecting force whatever the precise details of molecular interactions, it is appropriate to assume constant density arising from monomer-monomer hard core interactions. This assumption amounts to neglect the compressibility of DNA condensate inside the capsid. It is nevertheless possible to keep track of molecular interactions, and more generally of buffer conditions, which are well-documented with the help of the osmotic stress experiments on hexagonal phases of DNA \cite{rau_92,raspaud_05}, by allowing the constant density $\rho_0$ to depend on buffer conditions. We do not go further within this work in discussing its precise dependence, since our experiments were performed within roughly a single common buffer condition for both phages. Moreover the effect of ambiant salt on DNA ejection are addressed thoroughly elsewhere \cite{evilevitch_07}.

Using the aforementioned assumptions, the bending energy is calculated straightforwardly 
\begin{equation}
\label{energy bending}E_{bend}=-2\pi\rho_0d_0l_pkT\left[\sqrt{R_0^2-R^2}+R_0\ln\frac{R_0-\sqrt{R_0^2-R^2}}{R}\right]
\end{equation}
where the average inner radius of DNA condensate is $R$. Similarly, the DNA packaged length is rewritten
\begin{equation}
L_{in}(R)=\frac{4\pi\rho_0d_0}{3}(R_0^2-R^2)^{3/2}
\end{equation}

The ejecting force can be directly calculated as the derivative of this energy with respect to the length $L_{in}$. The result is
\begin{equation}
F(L_{in})=\frac{kTl_p}{2\left(R_0^2-\left(\frac{3L_{in}}{4\pi\rho_0d_0}\right)^{2/3}\right)}\equiv\frac{kTl_p}{2R^2}
\end{equation}

The use of constant density, together with the continuous approximation, leads to the artificial result of finite ejecting force $F_0=\frac{kTl_p}{2R_0^2}$ at zero loaded length, which represents the bending energy per unit length for a single hoop of radius $R_0$. This non-physical behavior can be corrected by substracting this constant residual force, such that the true ejecting force is now given by 

\begin{equation}
\label{force ejecting}F_{ej}(L_{in})=\frac{kTl_p}{2R_0^2}\frac{\left(\frac{3L_{in}}{4\pi\rho_0d_0}\right)^{2/3}}{R_0^2-\left(\frac{3L_{in}}{4\pi\rho_0d_0}\right)^{2/3}}
\end{equation}

In order to analyze the results of osmotic suppression datas, one needs to convert ejecting force into an outer osmotic pressure \cite{castelnovo_03,evilevitch_04}.  It has been shown that these experiments can be interpreted in terms of a force balance: the ejecting force acting on the base pairs located at the entry/exit of the capsid, which is due to the DNA remaining in the capsid, is balanced by the osmotic force resisting DNA insertion into the polymer solution. Notice that this osmotic force can be \textit{explicitely} associated to water molecule exchange between inside and outside the capsid (W.M. Gelbart, private communication). Up to the leading order, osmotic pressure and osmotic force are proportional $\Pi_{out}=\frac{4F_{osm}}{\pi d^2}\equiv\frac{4F_{ej}}{\pi d^2}$, with $d$ is the effective diameter associated to the insertion of DNA inside the polymer solution. In the case of poly-ethylene glycol (PEG), this diameter is estimated as the sum of DNA and PEG diameters, $d=2.4nm$ \cite{purohit_05,grayson_06}. Using this relation, Eq.\ref{force ejecting} is rewritten simply into two equivalent forms
\begin{eqnarray}
\label{reduced Pi}\tilde{\Pi}_{out} & = & \frac{\tilde{L}^{2/3}}{1-\tilde{L}^{2/3}}\\
\label{reduced L}\tilde{L} & = & \left(\frac{\tilde{\Pi}_{out}}{1+\tilde{\Pi}_{out}}\right)^{3/2}
\end{eqnarray}
with the following reduced variable $\tilde{L}=L_{in}/L_*$ and $\tilde{\Pi}_{out}=\Pi_{out}/\Pi_*$, and the characteristic parameters
\begin{eqnarray}
\label{l star}L_* & = & \frac{4\pi\rho_0d_0R_0^3}{3}\\
\label{Pi star}\Pi_* & = & \frac{2kTl_p}{\pi R_0^2d^2}
\end{eqnarray}

The equations \ref{reduced Pi} and \ref{reduced L} are the main theoretical results of the present work.
This analysis allows therefore to identify characteristic length scale and osmotic pressure for the problem of structureless DNA packaging. The equation \ref{reduced L} will be used in order to analyze osmotic suppression results described in next section. Then the value of characteristic parameters will be discussed extensively in section \ref{results}. 

Note that this model can be extended to different geometries of capsid, keeping the cylindrical symetry assumption for DNA condensate. In particular for a cylindrical capsid, a similar rescaling of experimental datas is predicted, cf Eq.\ref{reduced L}, where the only change is in the exponent $3/2$ that is replaced by 1, namely $\tilde{L}=\left(\frac{\tilde{\Pi}_{out}}{1+\tilde{\Pi}_{out}}\right)$.

\section{Results and discussion}
\label{results}

As it is described in the  section \ref{experiments}, the measurement of UV absorbance of digested nucleotides in the solution of phages in the presence of viral receptor allows to infer the amount of DNA ejected, and therefore the amount still present in the capsid. In the particular case of T5 phage, this is the first time the osmotic suppression technique is applied to monitor ejection of pressurized DNA out of the capsid, showing the applicability of this method to a different phage than $\lambda$, for which it has been originally designed \cite{evilevitch_03}. The results are shown in figure \ref{figure1}. The ejection from T5 phage (deletion mutant with 113kbp DNA) is completely suppressed at 7 atm osmotic pressure corresponding to 20$\%$ w/v of PEG8000. For comparison, ejection from $\lambda$ phage  (deletion mutant with 41.5kbp DNA) was completely suppressed at 15 atm corresponding to 30$\%$ w/v of PEG8000 \cite{evilevitch_03}. The relatively small value for outer osmotic pressure required to inhibit DNA ejection from the phage in the case of T5 have some important biological consequence, since it is believed that osmotic pressure in the cytoplasm of bacteria is of order few atmospheres (typically $\simeq 3atm$ ) \cite{serwer_88,neidhardt_96}. This means that full passive ejection of T5 genome in the bacteria is not possible, as ejection force is not strong enough to overcome the osmotic force resisting DNA insertion in the cytoplasm due to macromolecular crowding. Indeed, it is well-known that DNA transport into the host in the case of T5 is a two-step process, where 8$\%$ of the genome is transferred initially, allowing production of pre-early phage-encoded proteins that are necessary for transport of remaining DNA \cite{letellier_04,effantin_06}. Interestingly, our measurements indicate that for osmotic pressure of $3atm$, the ejected length represents roughly $9\%$ of wild-type genome (121.75kbp), in agreement with the \textit{in vivo} observation.

Previous kinetic measurements of T5 ejection triggered by FhuA receptor were described in references \cite{frutos_05a,frutos_05b,mangenot_05}. These experiments showed interesting stepwise ejections, suggesting particular features either in the structure of DNA itself, or in the conformation of DNA inside the capsid.
The main observation concerning osmotic suppression datas for T5 phage is that the partial suppression of DNA ejection seems progressive as function of polymer concentration in the phage solution. This is very similar to the case of $\lambda$ phage. Therefore DNA ejection out of T5 capsid, as monitored by osmotic suppression technique, does not allow to highlight the presence of steps in the ejection. Notice however that apparent continuous ejection in bulk experiments, where results are ensemble-average in the sense previously defined, is not inconsistent with the presence of steps in the ejection, as has been shown by De Frutos \textit{et al.} with light scattering technique \cite{frutos_05a}.

\begin{center}
\begin{figure}
\includegraphics[scale=0.6]{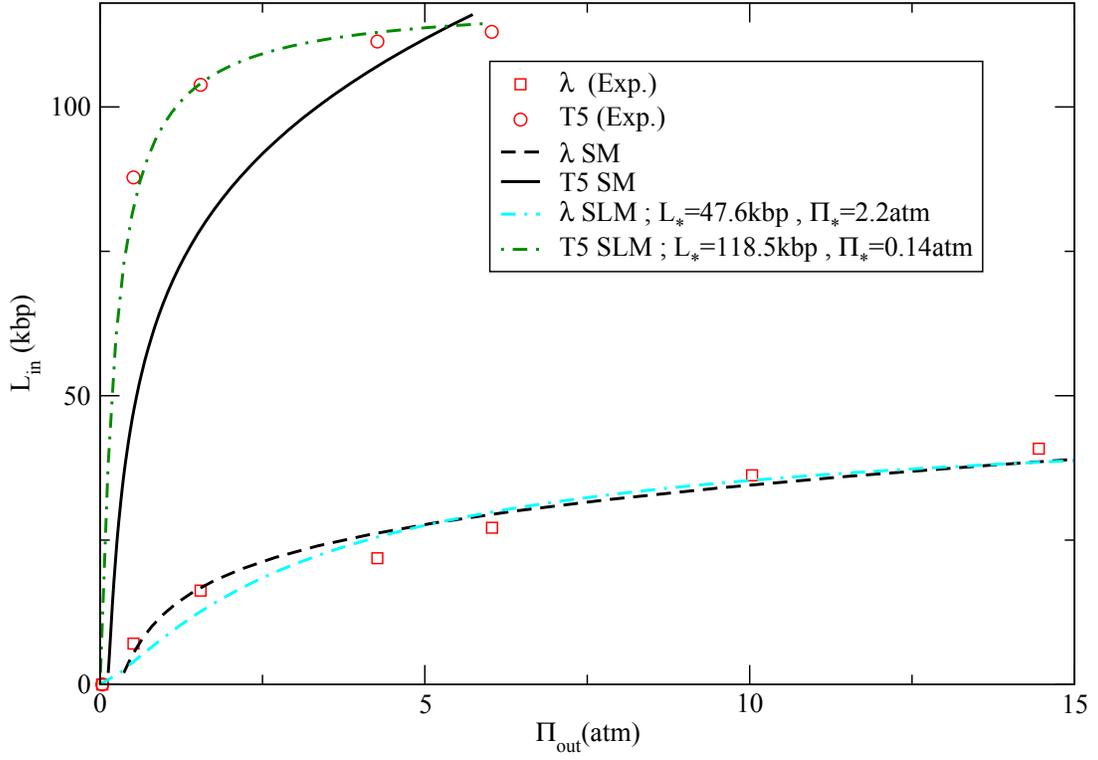}
\caption{\label{figure3}DNA length $L_{in}$ remaining inside $\lambda$ and T5 capsid as function of outer osmotic pressure $\Pi_{out}$. For $\lambda$, two buffer conditions are shown. $\lambda$\textit{+salt}\textit{SM}: structured model with the interaction parameters previously used to analyze $\lambda$ mutants \cite{grayson_06}. For the interstrand potential $V(d_s)=L_{in}\sqrt{3}F_0(c^2+cd_s)e^{-\frac{d_s}{c}}$, with $F_0=12000pN/nm^2$ and $c=0.3nm$. \textit{SLM}: structureless model, corresponding values of the fit parameters are shown.}
\end{figure}
\end{center}

The osmotic suppression datas for T5 and $\lambda$ phages are combined in figure \ref{figure3} for the sake of comparison. Notice that vertical axis is DNA length in kbp. The different amplitude of these curves reflects different nucleic acid content of the present strains, $L_{in}=113kbp$ and $L_{in}=41.5kbp$ for T5 and $\lambda$ respectively, as well as capsid internal radius, $R_0=39nm$ and $R_0=27.5nm$ for  T5 and $\lambda$ respectively \cite{effantin_06,dokland_93}. 
Despite differences in capsid sizes, DNA lengths and osmotic pressure required to suppress ejection, datas can be fitted using our structureless model (labeled as \textit{SLM} in figure \ref{figure3}) proposed in section \ref{model}. As can be seen by the comparison with previous structured models (labeled as \textit{SM} in figure \ref{figure3}) of $\lambda$ phage datas \cite{grayson_06}, the structureless model based only on bending consideration, is describing the experimental datas with similar accuracy. However, using the same $\lambda$ phage interstrand interaction parameters, namely $V(d_s)=L_{in}\sqrt{3}F_0(c^2+cd_s)e^{-\frac{d_s}{c}}$ with $F_0=12000pN/nm^2$ and $c=0.3nm$, in order to represent T5 datas,  there is a clear discrepancy between the structured model and experimental datas. This might be also attributed to slightly different buffer conditions used to perform experiments on T5 and $\lambda$, or to the presence of steps during the ejection from T5. The present set of measurements cannot discriminate between these scenarii. In the first hypothesis, however, we argue here that without any consideration on details of interactions arising from buffer conditions, the structureless model is nevertheless able to fit accurately T5 datas. Notice that minimum osmotic pressure required to inhibit completely DNA ejection from T5 phage ($\simeq 6-7atm$) is predicted accurately both by structure and structureless models, the latter being able to describe the whole data set.
The new information brought by the present analysis is the extraction of the two characteristic parameters $L_*$ and $\Pi_*$ for each phage.

First the value of characteristic length scale are $L_*=118.5kbp$ and $L_*=47.6kbp$ respectively for T5 and $\lambda$ phage. From these values, one can extract the corresponding uniform monomer densities for the two distinct phages using Eq.\ref{l star}. These informations can be converted to some typical effective diameter of DNA $d_0$ assuming local hexagonal order. The result is $d_0=2.67nm$ and $d_0=2.55nm$ respectively for T5 and $\lambda$. These values are consistent with values extracted from X-ray measurements for wild-type phages \cite{effantin_06,earnshaw_77}. It is also important to observe that in both cases $L_*$ are relatively close to the wild-type DNA content, $L_{T5wt}=121.75kbp$ and  $L_{\lambda wt}=48.5kbp$, although experiments were run with mutant phages with less nucleic acid content. Within this structureless model, the characteristic length scale is strictly associated to the divergence of ejection force as the curvature radius at the inner side of DNA condensate vanishes. The fact that wild-type DNA content in both cases are larger than this maximal theoretical threshold length of the model should be seen as the result of the use of constant density conformation in the calculation: different final conformations of packaged DNA within a spherical capsid lead to small fluctuations in bending energy, that were neglected in the model for the sake of calculation tractability. These fluctuations, that are difficult to quantify analytically, might be responsible for the shift in maximal DNA content of phage, as predicted by the characteristic length $L_*$. Within a more flexible interpretation of the model, this maximal length $L_*$ has to be related to some geometrical constraint of flexible tube packaging inside a spherical volume \cite{marenduzzo_03}. In other words, besides from bending or more general energetic considerations, it is reasonable to assume the existence of maximal \textit{geometrical} packaged length due to finite width of DNA.

The characteristic osmotic pressure extracted from measurements are $\Pi_*=0.14atm$ and $\Pi_*=2.2atm$ for T5 and $\lambda$ phage respectively. Using the definition Eq. \ref{Pi star} with the capsid radius $R_0=39nm$ for T5, the effective diameter $d=2.4nm$ of DNA when inserted in a PEG solution \cite{grayson_06}, and the persistence length of DNA $l_p=50nm$, the predicted value for $\Pi_{*T5}=0.14atm$ matches perfectly the experimental one. This is a strong point for the structureless model proposed here. However, the similar prediction for $\lambda$ phage ($R_0=27.5nm$) gives $\Pi_{*\lambda}=0.3atm$, a value significantly lower than the experimental one, $\Pi_*=2.2atm$. This discrepancy is likely associated to the neglected contribution of DNA condensate compressibility induced by curvature stress. Indeed, the characteristic osmotic pressure arising from the simple model can be interpreted as the only energetic scale of our system, as it is seen through its dependence with DNA bending modulus $\kappa=kT l_p$. Since the present model does not include the energetic contribution associated to DNA condensate compressibility, the difference between predicted and measured $\Pi_*$ can be considered therefore as an experimental measure of ensemble-averaged DNA-DNA interactions contribution. Another related source of possible discrepancy between the real system and the simple model is the assumed degree of ordering of DNA inside the condensate, which might be different for the two phages we are studying due to their different capsid structure and size: in the case of the well-ordered inverse spool conformation, the compressibility of DNA is expected be higher than for a disordered spool conformation, and therefore the energetic balance will be changed. Further development of the present model would be needed in order to address precisely these points, and this goes beyond the scope of this work.

\begin{center}
\begin{figure}
\includegraphics[scale=0.6]{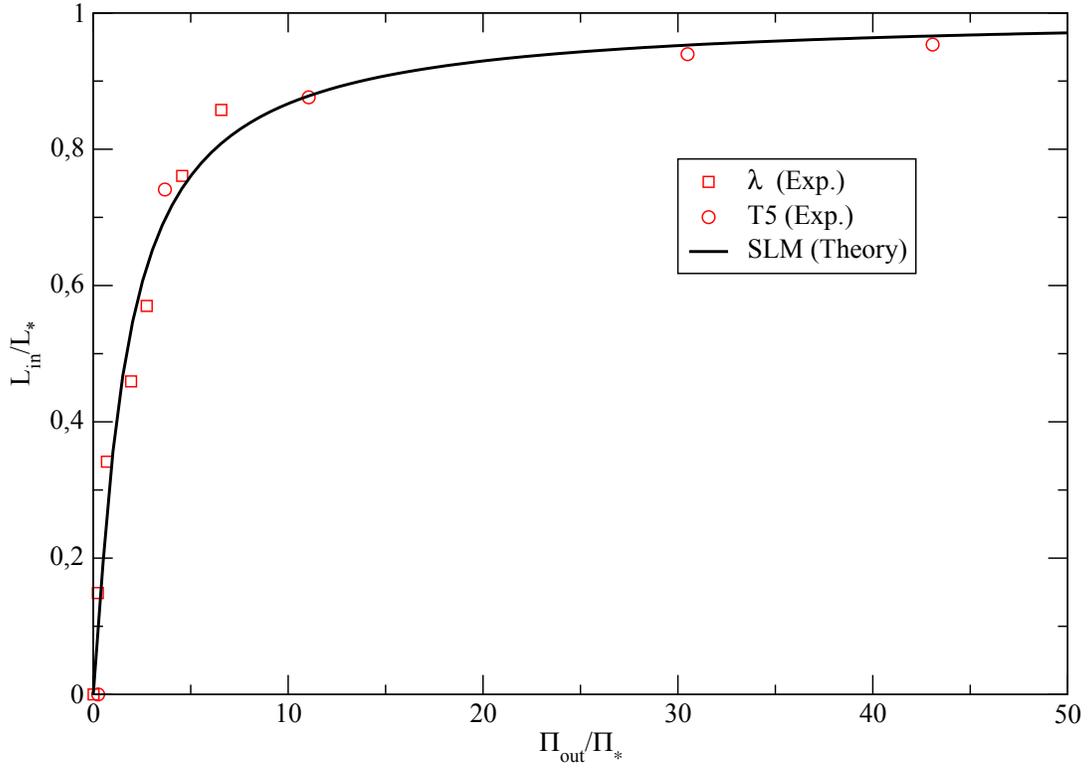}
\caption{\label{figure4}Rescaling of experimental variables ($L_{in},\Pi_{out}$) using the fit parameters ($L_*,\Pi_*$) for each phage. \textit{SLM}: structureless model prediction, cf Eq.\ref{reduced L}.}
\end{figure}
\end{center}

The results of the present analysis are conveniently represented by rescaling the experimental datas ($L_{in},\Pi_{out}$) for each phage by the characteristic parameters ($L_*,\Pi_*$). This is shown in figure \ref{figure4}. The result shows clearly that rescaled experimental datas fall on the same theoretical curve. Notice that since we used the extracted parameters to perform the rescaling, as opposed to the predicted parameters, this proves that identification of reduced equation \ref{reduced L} has a relevant meaning in analyzing the experimental datas, whatever the prediction for the characteristic parameters. Moreover the reduced nucleic acid length $\tilde{L}=L_{in}/L_*$ scales like the density of DNA inside the capsid $\rho$, or more roughly with the nominal nucleic acid volume fraction defined in section \ref{general}), up to some multiplicative constant. As it is  clearly observed in figure \ref{figure4} for the phage strains used in the present study, the density or volume fraction of nucleic acid inside both capsid is similar at full packaging. This might be related to some geometrical constraint of DNA packaging in spherical container. However, ejection force for T5 is lower than for $\lambda$ phage. This difference is an experimental proof that DNA packaging energetics is not simply determined by DNA density inside the capsid. The size of the capsids plays an important role as well.

Finally, let us mention that the present osmotic suppression datas obtained on T5 are not quantitatively inconsistent with an interpretation using the presence of kinetic stops during ejection. Indeed, in the case of stepwise ejection kinetics, one expects the underlying free energy describing DNA packaging (for example Eq.\ref{energy bending} in the case of the structureless model) to have one or several metastable minima. The datas are then interpreted using the lowest minimum of the free energy while increasing osmotic pressure. In the case of continuous ejection, \textit{i.e.} lowest free energy minimum is full ejection at zero osmotic pressure, the equation of state $L_{in}\, vs\, \Pi_{out}$ is monotically increasing. In the case of a single secondary  metastable minimum, a DNA length jump should occur at some osmotic pressure between the two distinct minima (data not shown for the sake of brevity). In other words, for this precise osmotic pressure, two different amounts of DNA, corresponding to equal packaging free energies, will be found in the phages of the solution. The net shape of the equation of state is characterized by a vertical jump at the threshold osmotic pressure. In the case of our T5 datas, we cannot exclude the presence of such a jump at very low osmotic pressure ($\simeq 1-2atm$). However, precise modelization of stepwise features in this context goes beyond the scope of the present work, which is mainly focused on the similarities of osmotic suppression datas on different phages.

\section{Concluding remarks}
\label{conclusion}
In this work we presented osmotic suppression measurements on T5 and $\lambda$ phages, therefore extending the bearing of osmotic suppression technique to different phages. In the particular case of T5 phage, ejection looks very similar to $\lambda$ through the osmotic suppression technique, but presence of stops during the ejection kinetics cannot strictly be excluded. The amplitude of measured ejecting force is also quantitatively consistent with the two-steps mechanism of ejection reported in the litterature \cite{letellier_04,effantin_06}. The first step of ejection is thought to be spontaneous after receptor binding (passive ejection due to energy stored in the DNA condensate), releasing 8$\%$ of wild-type T5 genome. In the second step, the remaining DNA is transfered to the host by a protein-dependent translocation mechanism. According to our \textit{in vitro} experimental results, passive ejection is limited to 9$\%$ of wild-type T5 DNA in the presence of stressing agents, which are mimicking in a quantitative way the  osmotic pressure of cell cytoplasm ($\simeq 3atm$).

The experimental results were additionally analyzed within the framework of a simplified model based on bending considerations only. The major assumption of this model is that leading contribution to the force ejecting nucleic acid out of bacteriophage, is due to the bending penalty of confining DNA. As compared to previous modeling that include semi-empirical DNA-DNA interstrand interactions in a well-ordered conformation as well, the datas on T5 and $\lambda$ are well described with similar accuracy. In other words, the length dependence of ejecting force is accurately described by a structureless model of DNA condensate with uniform density. Given the recent simulation results that did highlight the quenched fluctuations in final conformations of DNA within virions \cite{muthukumar_06,locker_06,petrov_07}, and therefore raised some fundamental points in the assumption of structured models, the present simple model might provide a good alternative to interpret ensemble-averaged experimental results like osmotic suppression measurements.
The theoretical analysis allowed to identify a reduced equation, Eq. \ref{reduced L}, that allows to rescale results of osmotic suppression datas for T5 and $\lambda$ phages.

At this stage, more analytical or numerical work is still to be done in order to elucidate the ensemble-average behavior for the different final conformations, and in particular the relative weight of bending and short-range interaction contributions to the net force ejecting DNA from bacteriophage. This balance might depend on the size of the capsid as well, as simulation  results on $\phi 29$ phage packaging \cite{locker_06,petrov_07} seems to indicate that the very last steps of DNA packaging in this particular virus, which has a smaller capsid (average radius $R_{\phi 29}= 25nm$) than T5 and $\lambda$ phages, are dominated by electrostatic interactions, although the innermost DNA region is disordered.

More specifically, the possibility to compare experimentally DNA ejection force for different phages within a single technique represents a significant step towards the understanding of the relation between capsid size, DNA content and the strength of capsids. The capsid size (inner radii $R_{T5}=39nm$ and $R_{\lambda}=27.5nm$) and nucleic acid content ($L_{T5}=121.75kbp$ and $L_{\lambda}=48.5kbp$) are both significantly different for wild-type T5 and $\lambda$ bacteriophages. However, the nominal volume fraction of nucleic acid (see section \ref{general} for definition) is similar for both phages ($\phi_{T5}=0.52$ and $\phi_{\lambda}=0.59$). This fact is likely to be related to some geometrical constraint associated to the packaging of DNA inside the capsid. Within the model proposed in this work, this common density is correlated to the existence of the characteristic length $L_*$ that rescales the nucleic acid content $L_{in}$ for each phage. If the DNA volume fraction inside the capsid is the only relevant variable in equilibrium DNA packaging energetics, one might expect ejection force to be similar for T5 and $\lambda$. Our experimental results proves clearly that this is not the case. Therefore at least one more relevant variable is necessary to describe equilibrium DNA packaging energetics. In the framework of the simple model derived in this work, this variable is the characteristic ejection force amplitude $F_{*}=\frac{kTl_p}{2R_0^2}$, which can be modulated either by DNA persistence length $l_p$ or the capsid size $R_0$. According to this model, smaller phages are typically expected to require higher osmotic pressure in order to inhibit completely the ejection of their DNA.

The difference in ejecting force for fully packaged bacteriophage strains used in this work ($F_{T5}\simeq 3pN$ and $F_{\lambda}\simeq 7pN$) are likely to reflect different maximal mechanical strength of capsids, and also different mechanism of DNA transport into the host. In the case of $\lambda$ phage, the build-up of large internal stress upon packaging as compared to T5 phage requires a stronger capsid. Similarly, the large internal stress allows $\lambda$ phage to rely mostly on passive ejection for DNA to be transported into the host. In the case of T5 phage, the genome is longer and more complex than $\lambda$. Its large capsid size allows it nevertheless to have less elastic energy stored within DNA condensate. This lower energy is compensated by the larger genetic information encoded into T5 DNA that allows it to have both passive and ``active'' DNA transport into the host.

\begin{acknowledgments}
We greatly acknowledge Lucienne Letellier and Marta de Frutos for providing us with bacteriophage T5 and FhuA receptor. We want to thank William Gelbart and Charles Knobler for much valuable advice to improve this work. Useful suggestion of a referee about the simpe model is also acknowledged.
We thank Paulo Tavares for discussions. This work was supported by the Swedish Research Council (VR) and Crafoord Foundation through grants to AE.
\end{acknowledgments}

\bibliography{biblio}

\newpage

\end{document}